\documentclass[useAMS,usenatbib]{mn2e}
\usepackage{graphicx}
\title[Variability Among Stars in the M\,67 Field]{Variability Among Stars in the M\,67 Field from {\it Kepler/K2-Campaign-5} Light Curves}
\author[G. Gonzalez]{Guillermo Gonzalez$^{1}$\thanks{E-mail:
ggonzalez@bsu.edu}\\
$^{1}$Ball State University, Department of Physics and Astronomy, 2000 W. University Ave., Muncie, 
IN 47306}
\begin{document}

\date{Accepted ??. Received ??; in original form ??}

\pagerange{\pageref{firstpage}--\pageref{lastpage}} \pubyear{??}

\maketitle

\label{firstpage}

\begin{abstract}
We examine the photometric variability of stars in the M\,67 field using {\it Kepler/K2-Campaign-5} light curves. Variabilities and periods were determined for 639 stars. The mean photometric period of 28 single Sun-like members stars in M\,67 is $23.4 \pm 1.2$ d. This corresponds to a gyro-age of $3.7 \pm 0.3$ Gyr, assuming the periods can be associated with rotation. The intrinsic variabilities of the solar analogs are greater than the Sun's variability, as measure from VIRGO fluxes. We also find evidence that the single cluster members have a different distribution of variability than the binary members.
\end{abstract}


\section{Introduction}

The open cluster M\,67 has several attractive qualities that make it well-suited for exploring a number of current topics in astrophysics and astrobiology. Its composition is very similar to the Sun's \citep{onehag14}; it has small and uniform  interstellar extinction \citep{tay07}; it is one of the nearest clusters, at about 850 pc \citep{yak09}; it is nearly the same age as the Sun \citep{sara09,yad08}. M\,67 is also an important testing ground for new stellar isochones \citep{mow12} and new physics in stellar interiors \citep{mag10}. It is also one of the few clusters in which planets have been detected \citep{bru14}.

Membership probabilities are available from proper motion measurements for thousands of stars in the M\,67 field from multiple sources \citep{sand77,gir89,yad08,nard16}, as well as from radial velocities \citep{gell15}. These permit the creation of a color-magnitude diagram (CMD) cleaned of field nonmembers, as well as reliable separation of the single and binary star cluster members.

With a Galactic latitude of nearly 32 degrees, the M\,67 field is sparse, making it a particularly attractive target for telescopes optimized for high photometric precision rather than high angular resolution. One such instrument is the NASA {\it Kepler} space observatory. Each pixel in {\it Kepler's} CCDs subtends an angle of 4 arc seconds on the sky. Given this, contamination with light from unresolved faint sources is a concern in crowded fields observed with {\it Kepler}, such as the original main mission field. Aperture photometry of stars in the M\,67 field, however, should be less affected by contamination. 

The {\it Kepler} main mission ended in 2013, and {\it Kepler 2.0/K2} began in 2014. The data collected and analyzed to date demonstrate that the photometric precision achieved with {\it K2} for stars fainter than 12 magnitude is close to that achieved during the main mission \citep{aigrain15}. The M\,67 field was selected for inclusion in {\it Kepler/K2-Campaign-5}, which took place in spring 2015.

M\,67 has a number of well-studied high-amplitude variables, such as HV Cnc and EU Cnc. In addition, low-amplitude variability has been detected in several dozen other cluster members \citep{stass02,nard16}. Old sun-like stars typically vary by a few millimagnitudes (mmag) over several years, while some vary less than 1 mmag \citep{lock13}. While the short time duration of the {\it Kepler/K2} observations of M\,67 prevents an analysis of long-term Sun-like activity cycles amongst its member stars, it should serve as an excellent dataset for analysis of short-term (hours to weeks) variability.

The purpose of the present work is to explore the photometric variability of M\,67 member stars using the {\it Kepler/K2 Campaign-5} data. We describe and prepare the data for analysis in Section 2. In Section 3 we discuss the data analysis, while in Section 4 we discuss the results. We present our conclusions in Section 5.

\section{Data Preparation}

M\,67 was observed continuously between April 27 and July 10, 2015 during the {\it Kepler/K2-Campaign-5}  (hereafter, 'Campaign-5 field'). It includes 28,850 long-cadence, 204 short cadence, and several other special targets. Several data products for the Campaign-5 field were released to the public on the NASA Barbara A. Mikulski Archive for Space Telescopes (MAST) website on October 30, 2015.\footnote{https://archive.stsci.edu/k2/} We downloaded tar files containing all the long cadence light curve (CLC) files of the Campaign-5 targets from the MAST archive. In addition, we downloaded the comprehensive {\it K2} input catalog (EPIC) for the Campaign-5 field.

We supplemented the NASA {\it K2} data with ground-based data, of which \citet{nard16} is our primary source. \citet{nard16} list the positions and white-light magnitudes for 6905 objects in the M\,67 field, but they only list $BVRI$ magnitudes, proper motions and membership probabilities for a subset of this large sample. Cross-referencing (using coordinates) the Campaign-5 input catalog with the \citet{nard16} catalog resulted in 3201 matches. Of these, 639 have light curves available in the MAST Campaign-5 archive. This will be the working sample.

We plot the locations of our working sample stars in the M\,67 field in Figure 1. The outer perimeter of the distribution of the stars in the figure are determined by the field of view of the \citet{nard16} images (see their Figure 2). Most of the stars in our sample are more than a quarter of a degree from the cluster center. However, M\,67 is often described as subtending an angle of half a degree on the sky. Thus, most of our sample stars are in the outskirts of the cluster. The number of working sample stars is sparse in the inner region of the cluster because we did not include in our analysis the special 400x400 pixel region centered on the cluster center.

\begin{figure}
\includegraphics[width=3.5in]{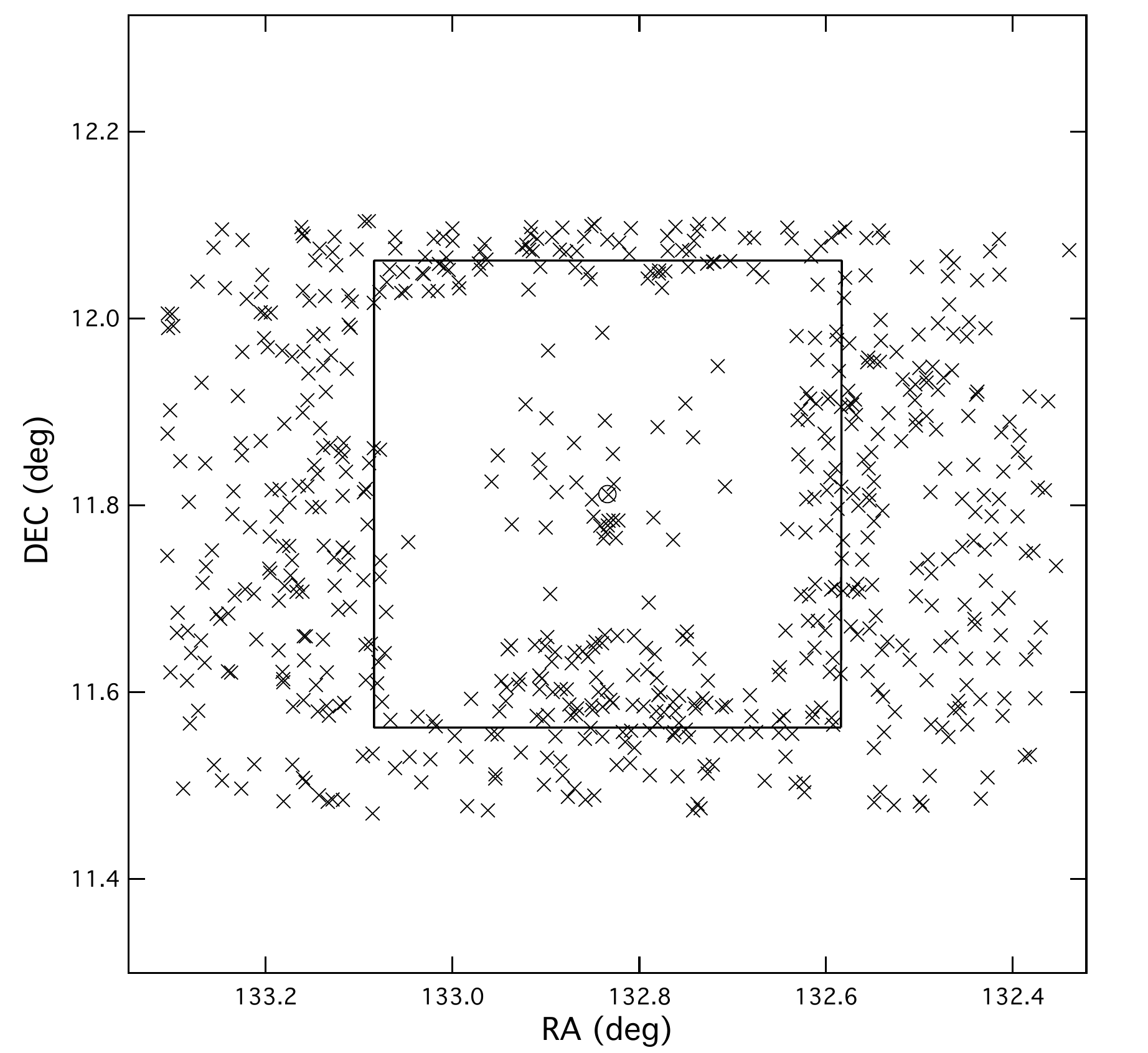}
\caption{Working sample stars in the field of M\,67. The center of the cluster is marked with an open circle. The square spans half a degree on a side. North is up and east is to the left.}
\end{figure}

The stars in our working sample range in $V$ from 8.9 to 21.4 magnitude. Nearly all stars in the M\,67 field brighter than $V = 15$ have published proper motion membership probabilities \citep{sand77,gir89,yad08,nard16}, and many also have radial velocity membership probabilities \citep{gell15}. All but 39 stars in our sample have proper motion membership probabilities from \citet{nard16}. We have cross-matched the catalogs of \citet{gell15} and \citet{yad08} with our sample in order to fill-in missing membership probabilities as well as to check their consistency when multiple values are available. 

Following \citet{gell15}, we also assigned a ``membership class'' to each star (see their Table 4). The membership class was straightforward to assign in most cases. In a few cases we changed the class they assigned, mostly from ``U'' (unknown membership) to ``M'' (member) or ``SN'' (single non-member) to ``SM'' (single member). We also changed each instance of ``BLM'' (binary likely member) and ``BLN'' (binary likely non-member) to ``BM'' or ``BN'', as the proper motion data warranted. The classes ``SM'', ``BM'', ``SN'', and ``BN'' can only be assigned to those stars with radial velocity data. In cases where only proper motion membership probabilities are available, we assigned class ``N'' (non-member) to stars with probability less than 20\% and class ``M'' (member) to stars with probability greater than 80\%. We list in Table 1 the tally of stars from our sample for each class. For the remaining stars we assigned class ``U'' (unknown). Nearly all the stars with the ``U'' class are fainter than $V = 19$ magnitude.

\begin{table}
\caption{Number of stars in each class in our working sample}
\label{obs}
\begin{tabular}{@{}llc}
\hline
Class & Description & Number \\

\hline
M & member & 105 \\
SM & single member & 130 \\
BM & binary member & 37 \\
N & non-member & 106 \\
SN & single non-member & 95 \\
BN & binary non-member & 18 \\
U & unknown membership & 148 \\
\hline
\end{tabular}

\end{table}

The {\it K2} mission is designed to observe fields along the ecliptic plane. However, due to the lack of precise pointing control, the {\it Kepler} telescope slowly drifts as it collects data, requiring corrective telescope moves about every 6.5 hours to keep the targets of a particular campaign within the field-of-view. These pointing corrections result in sudden and large changes in the star positions on the CCDs, which, in turn, cause large jumps in the observed target fluxes. Data collected during telescope moves have been assigned a value greater than zero for the \texttt{SAP\_QUALITY} flag for each light curve; only data collected with values of zero for this flag were retained for further analysis in the present work. 

We used the \texttt{PDCSAP\_FLUX} values for our analysis of each light curve. As described in the Kepler Archive Manual,\footnote{https://archive.stsci.edu/kepler/manuals/} dated June 5, 2014, these aperture flux values have been corrected for systematic instrumental trends using a Bayesian statistical approach. Visual inspection of several light curves confirmed that the use of these flux values (with corresponding \texttt{SAP\_QUALITY} flag value of zero) eliminated nearly all the obvious outliers and the systematic trends.

Next, we calculated the mean flux, the mean estimated flux error (from \texttt{PDCSAP\_FLUX\_ERR}), and the flux variance using the high quality measurements (\texttt{SAP\_QUALITY} flag $= 0$) for each light curve in our working sample. Visual examination of the data revealed some obvious outliers in a few of the light curves. In order to eliminate these few discrepant measurements, we deleted flux values that deviated by more than 4 sigma from the mean flux. Note, this procedure not only eliminates truly discrepant flux values due to instrumental systematics, but it also removes extreme flux values due to flares and deep eclipses; this does not pose a problem for the present study, however, since we are interested only in slow activity-related brightness variations. Finally, we calculated a new set of flux mean and variance values to replace the original ones.

We have also calculated a quantity we call the ``flux variability index'' (FVI), which is simply the ratio of the flux standard deviation to the mean estimated flux error; if this ratio is significantly greater than one, then there must be an additional source of variability not included in the estimated flux error, which is presumably astrophysical in origin. The FVI values range from 1.1 to 838. The FVI values will be our primary test for variability among our sample stars. The fact that the smallest FVI values are only 10\% greater than unity implies that the estimated photometric errors are an accurate measure of the actual errors and that systematic errors have been adequately accounted for.

We show in Figure 2 both the estimated flux error and measured standard deviation for the stars in our working sample; four stars were left out of the plot because they lack $R$ magnitudes. There is an obvious gap between the flux error and the standard deviation for stars brighter than about $R = 18$. If we assume that the systematic errors have been properly removed, then the gap implies that the intrinsic variability has been clearly detected for all stars brighter than 18 magnitude and some of the fainter ones. We plot the FVI values in Figure 3. 

\begin{figure}
\includegraphics[width=3.5in]{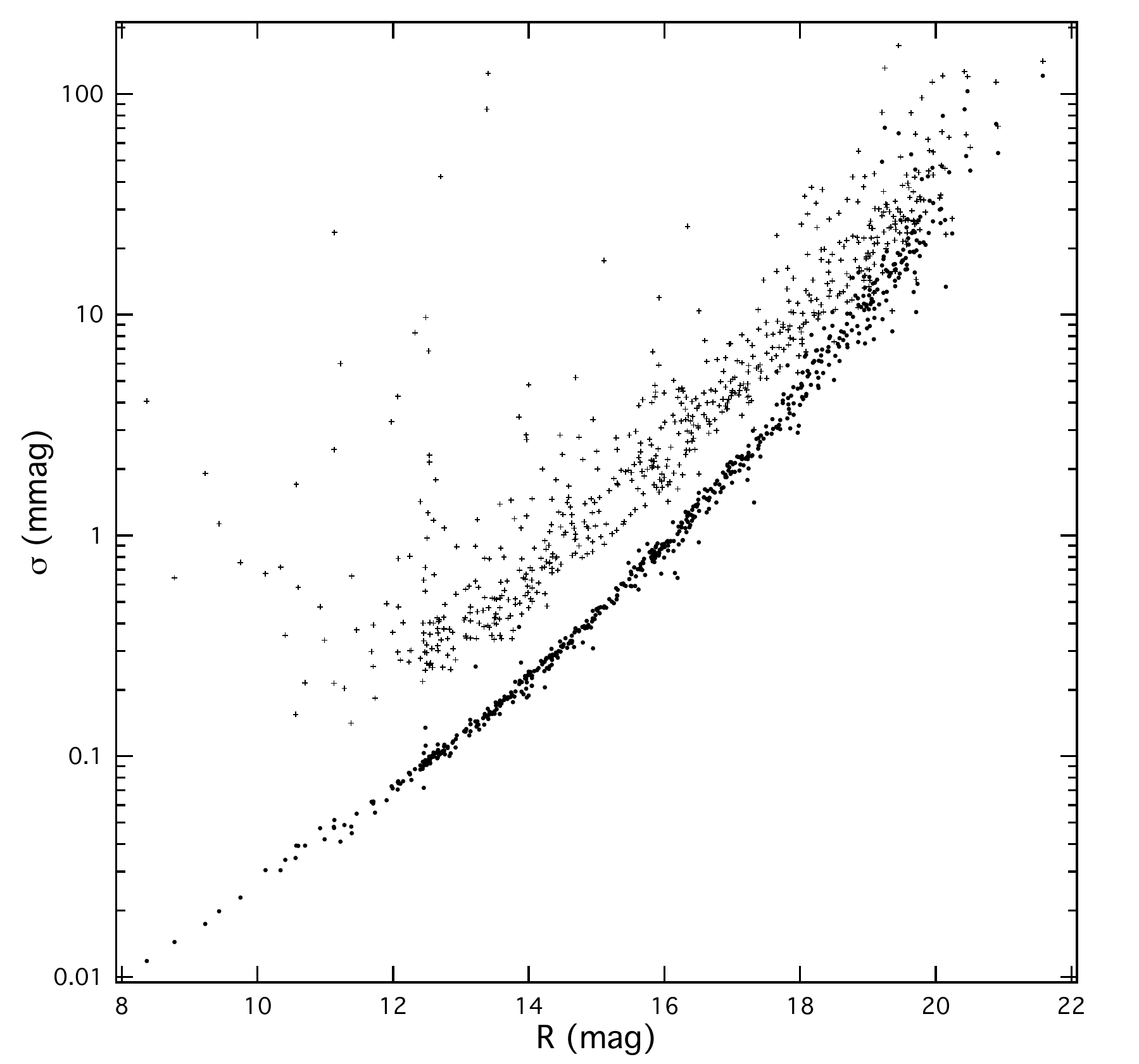}
\caption{Estimated flux error (dots) and measured standard deviation (plus signs) are plotted against $R$ magnitude from \citet{nard16} for our working sample.}
\end{figure}

\begin{figure}
\includegraphics[width=3.5in]{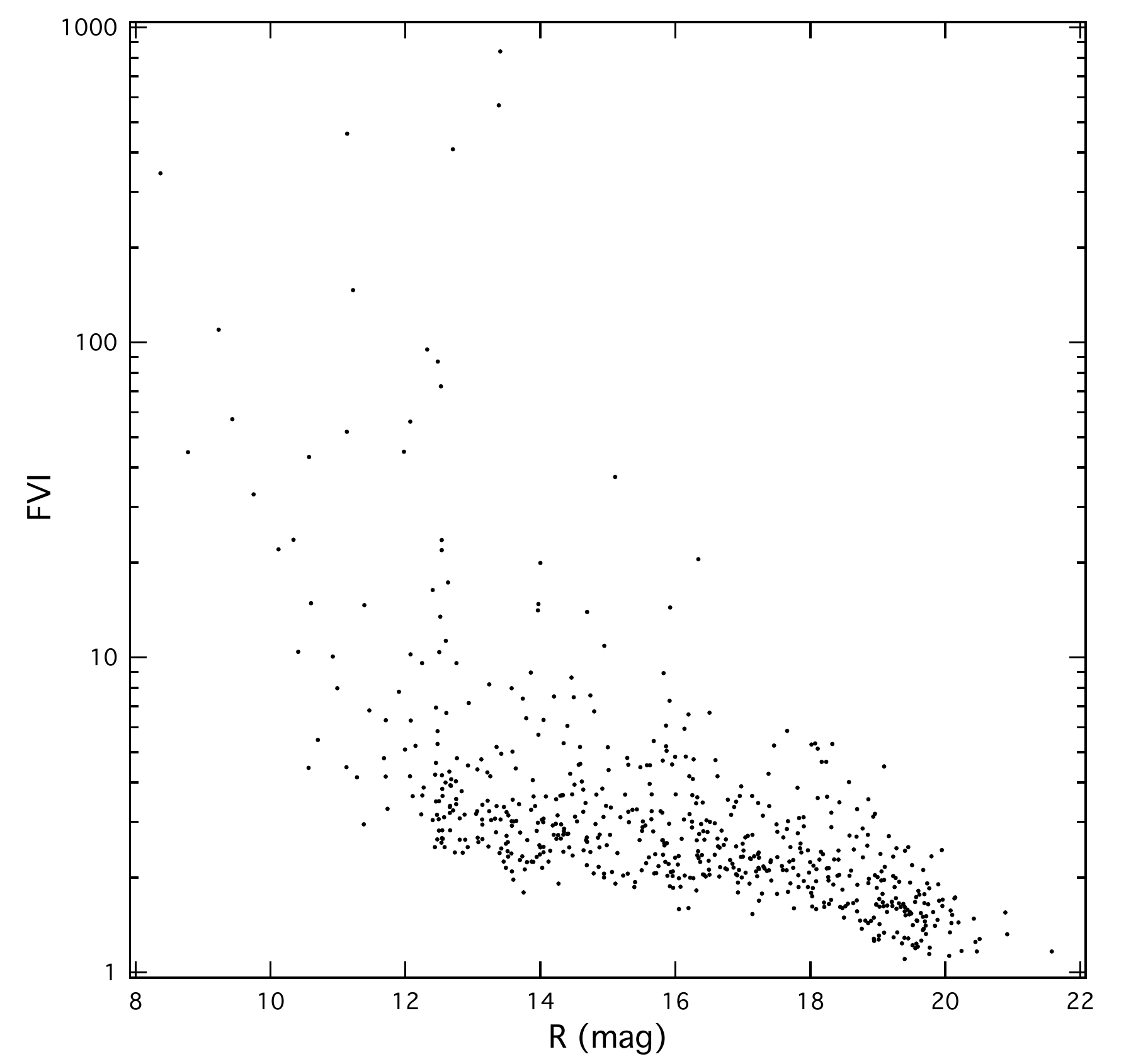}
\caption{FVI values are plotted against $R$ magnitude for our working sample.}
\end{figure}

In Figure 4 we plot the distribution of raw variabilities (square root of the variance converted to millimagnitudes, mmag) for all the stars in our working sample. Raw variabilities include measurement errors, unmodeled systematic error, and intrinsic astrophysical variability. The values range from 0.14 to 235 mmag. In addition, we calculated a second ``corrected'' variability for each star by subtracting the estimated error variance from the measured flux variance and then converting to mmag. The largest changes occurred for stars with the largest variabilities; the largest variability was reduced to 168 mmag. There is a sharp drop-off in the number of stars with variability values below 0.2 mmag; the smallest variability value was reduced to 0.13 mmag.

\begin{figure}
\includegraphics[width=3.5in]{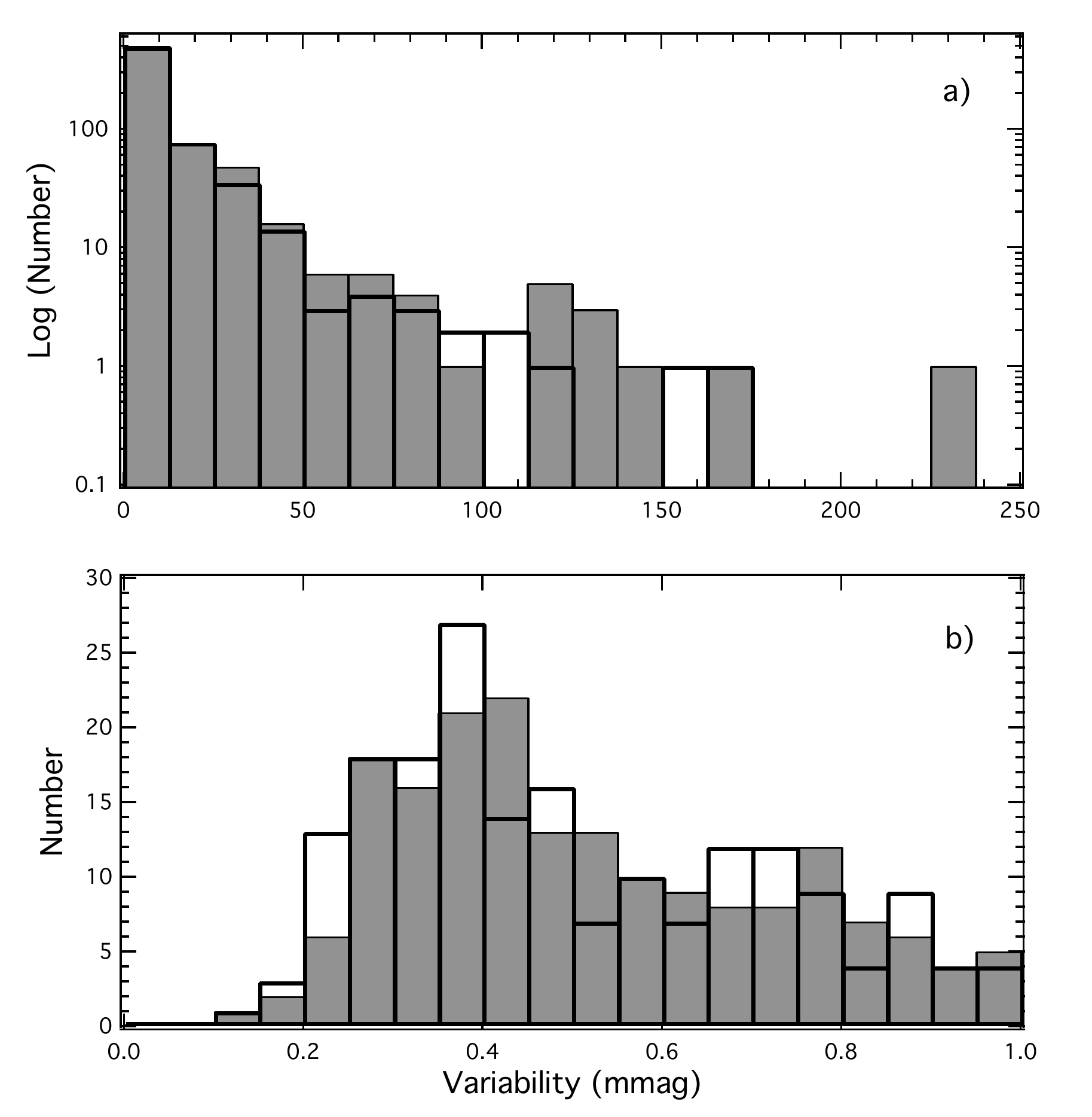}
\caption{Distribution of raw variability values for the stars in our working sample are shown (filled bars) over the full range of variability (panel {\bf a}) and for variability values less than 1.0 mmag (panel {\bf b}). The corrected variability values are shown as unfilled bars in both panels. See text for details.}
\end{figure}

Another quantity we want to determine for each star is the rotation period. Starspots regularly rotating into and out of view cause photometric variations that have been measured for many Sun-like stars. \citet{mcq14} measured the rotation periods of over 34,000 stars in the original {\it Kepler} field. To efficiently determine rotation periods for the stars in our working sample, we employed a Lomb-Scargle periodogram analysis method implemented in Python called gatspy; the actual function is called ``LombScargleFast.''\footnote{http://www.astroml.org/gatspy/} For each star we visually examined its periodogram. We also used gatspy to automatically search for the optimum period in the range 0.05 to 40 days using a dense frequency grid and plotted a phased light curve using the optimum value of the period. 

Based on the appearances of the periodogram and light curve for each star we assigned one of the following subjective categories: ``rot'' (rotation modulated), ``mult'' (multiple periods), ``EB'' (eclipsing binary), ``none'' (no clear variation). If the periodogram is dominated by one peak and the light curve shows a pattern suggestive of rotational modulation, we assigned it to the ``rot'' category. If the periodogram has two or more peaks of comparable power and the light curve appears to have multiple periods, then we assigned it to the ``mult'' category. If the periodogram peaks are weak and the light curve lacks a discernible pattern, we assigned it to the ``none'' category. We list these and other data for each star in our working sample in Table 2 (full version available online).

We plot sample periodograms of stars in the ``EB'' ,``mult'' and ``none'' categories in Figure 5; samples in the ``rot''  category are shown in the next section. For EPIC211408138, the derived optimum period is 5.17 d, which coincides with the sharp high peak in the top panel of the figure. Examination of its phased light curve indicated that this is the correct (orbital) period for this star. In the case of EPIC211427909, the gatspy-derived optimum period is 15.48 d, which corresponds to the highest peak in the middle panel. We placed this star in the ``mult'' category because the two highest peaks in the periodogram are very close in height and the light curve phased according the optimum period has multiple superposed variations. Lastly, the optimum period for EPIC211399673 is 0.999 d. We placed it in the ``none'' category, because it exhibits very weak peaks in the periodogram, and the phased light curve does not show an obvious pattern. We should note that we assigned periods to all the stars in our working sample, regardless of the category.

\begin{figure}
\includegraphics[width=3.5in]{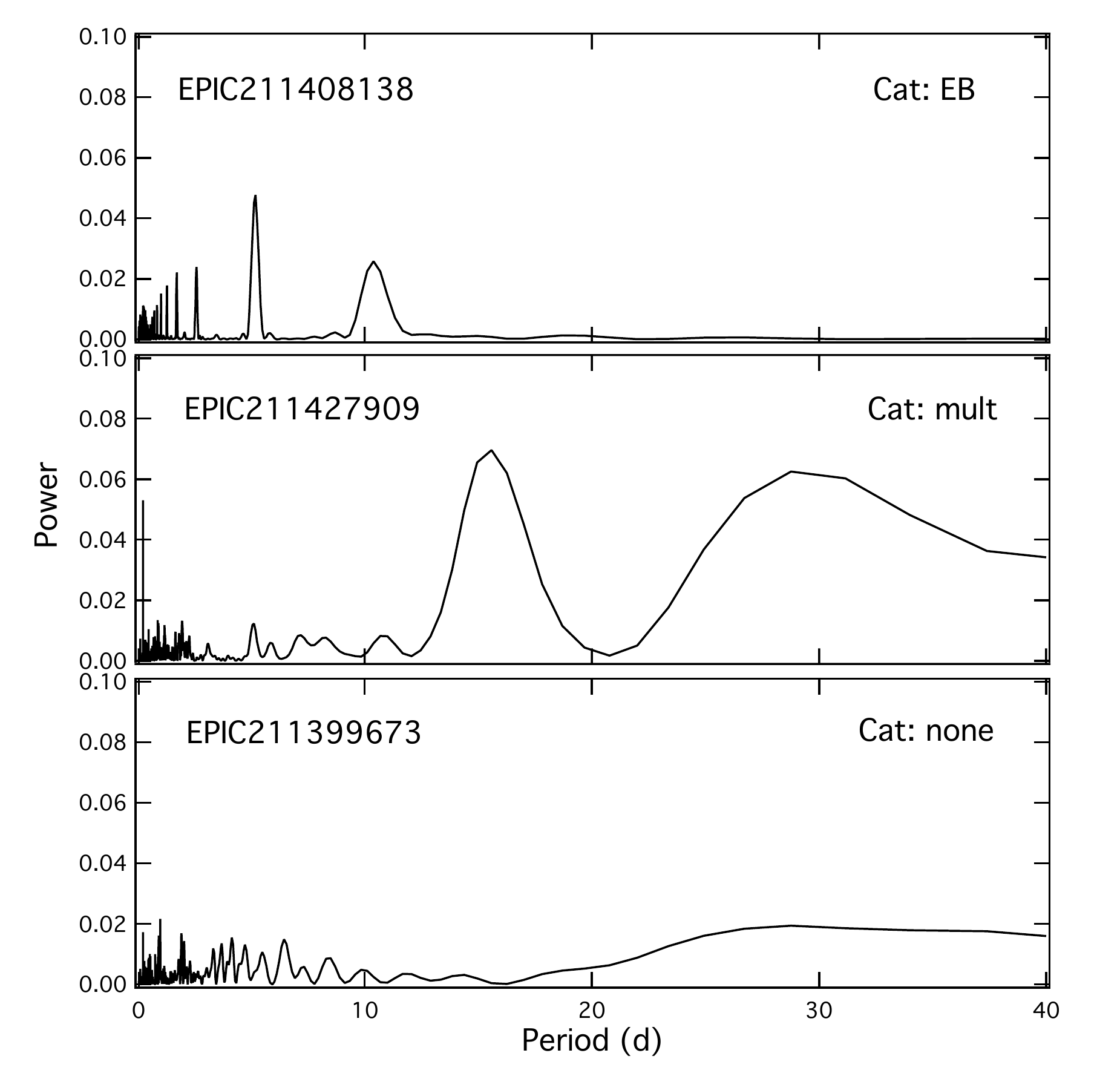}
\caption{Sample periodograms for stars in three of our four categories. See text for details.}
\end{figure}

\begin{table*}
\centering
\begin{minipage}{160mm}
\caption{Various data for the stars in our working sample. The complete table is available as on online supplement.}
\label{xmm}
\begin{tabular}{lrcccrcccc}
\hline
EPIC & Other ID & Class & Category & Period & FVI & Variability & Corrected variability & RA (2000.0) & DEC\\
 & & & & (d) & & (mmag) & (mmag) & (deg) & (deg) \\
\hline
211389202 & & N & rot & 27.949959 & 2.44 & 11.258 & 10.268 & 133.085563 & 11.470046\\
211389426 & & N & mult & 35.482820 & 5.31 & 36.910 & 36.249 & 132.961953 & 11.473468\\
211389428 & & M & rot & 28.722914 & 2.32 & 4.994 & 4.509 & 132.742179 & 11.473551\\
\hline
\end{tabular}
\end{minipage}
\end{table*}

Although the M\,67 field is sparse and crowding is minimal, there might be light contamination from other stars inside or near the apertures of some of the target stars. In order to check on the possible importance of such light contamination on variability, we have calculated for each star in our working sample the angular distance to the closest star in the full \citet{nard16} catalog (starting at a minimum separation of two arc seconds, which is just a bit larger than the two-pixel resolution of their images). If light contamination is an important source of flux variability, then we should detect a dependence of the amount of variability on the angular separation from the closest star \citet{welch93}. We did not find any evidence for such a correlation.

In the next section we will compare the variability of Sun-like stars in M\,67 and the Sun. In preparing a suitable set of comparison solar data, we follow a procedure similar to that of \citet{basri13}. In particular, we make use of the VIRGO total solar irradiance data from the {\it Solar and Heliospheric Observatory} (SOHO) spacecraft. It is a nearly ideal dataset to compare to {\it Kepler} photometry, given its similar spectral response; see \citet{fro01} for additional details. We downloaded the hourly VIRGO data,\footnote{ftp://ftp.pmodwrc.ch/pub/data/irradiance/virgo/TSI/virgo\_tsi\_h\_v6\_004\_1510.dat} which runs from January 1996 to October 2015. This range covers all of Solar Cycle 23 and part of 24. Thus, it samples the full range from the quiet to the active Sun.

In order to fairly compare the solar irradiance variations to the {\it Kepler} photometry, we applied the following edits to the VIRGO data. First, we extracted at random a section of the VIRGO data covering the same time interval as the {\it K2} M\,67 data. If the data section contained more than four consecutive missing data rows, it was discarded and a different section of data was selected, and the process was repeated until a sufficiently complete data section was found. Next, the missing data were filled in using linear interpolation. Then, the VIRGO hourly measurements were interpolated to match the {\it K2} half-hour cadence. Next, the VIRGO data were matched to the \texttt{SAP\_QUALITY} flag values taken from the {\it K2} dataset, and, just as with our procedure described above, we retained for analysis only the data with quality values equal to zero. Next, we calculated the mean and variance of the flux and deleted flux values more than 4 sigma from the mean and then recalculated the mean and variance. Finally, the period was determined from the periodogram analysis. This entire procedure was repeated 1000 times.

We show in Figure 6 the distribution of the derived solar rotation periods. The mean period is $25.2 \pm 8.1$ days, the median is 25.7 days, and the mode is 28 days. These are close to the synodic Carrington solar rotation period of 27.3 days (the corresponding sidereal period is 25.4 days). We show in Figure 7 the distribution of the resulting solar photometric variability (= measured standard deviation) converted to mmag units. The mean, median and mode are $0.20 \pm 0.14$,  0.18 and 0.06 mmag, respectively. 

\begin{figure}
\includegraphics[width=3.5in]{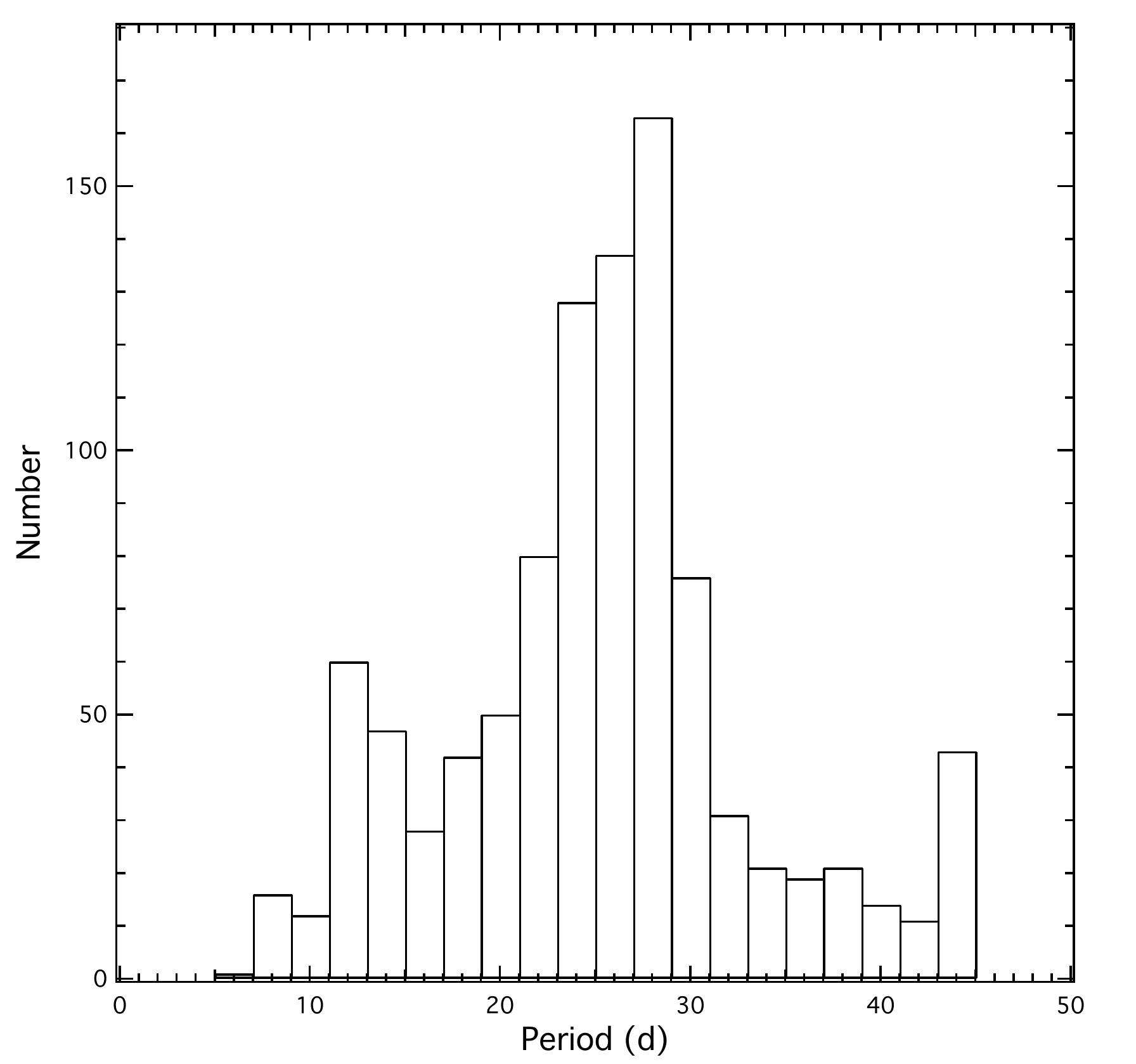}
\caption{Distribution of solar rotation periods obtained from periodogram analyses of 1000 sample photometric datasets extracted from the complete VIRGO data set.}
\end{figure}

\begin{figure}
\includegraphics[width=3.5in]{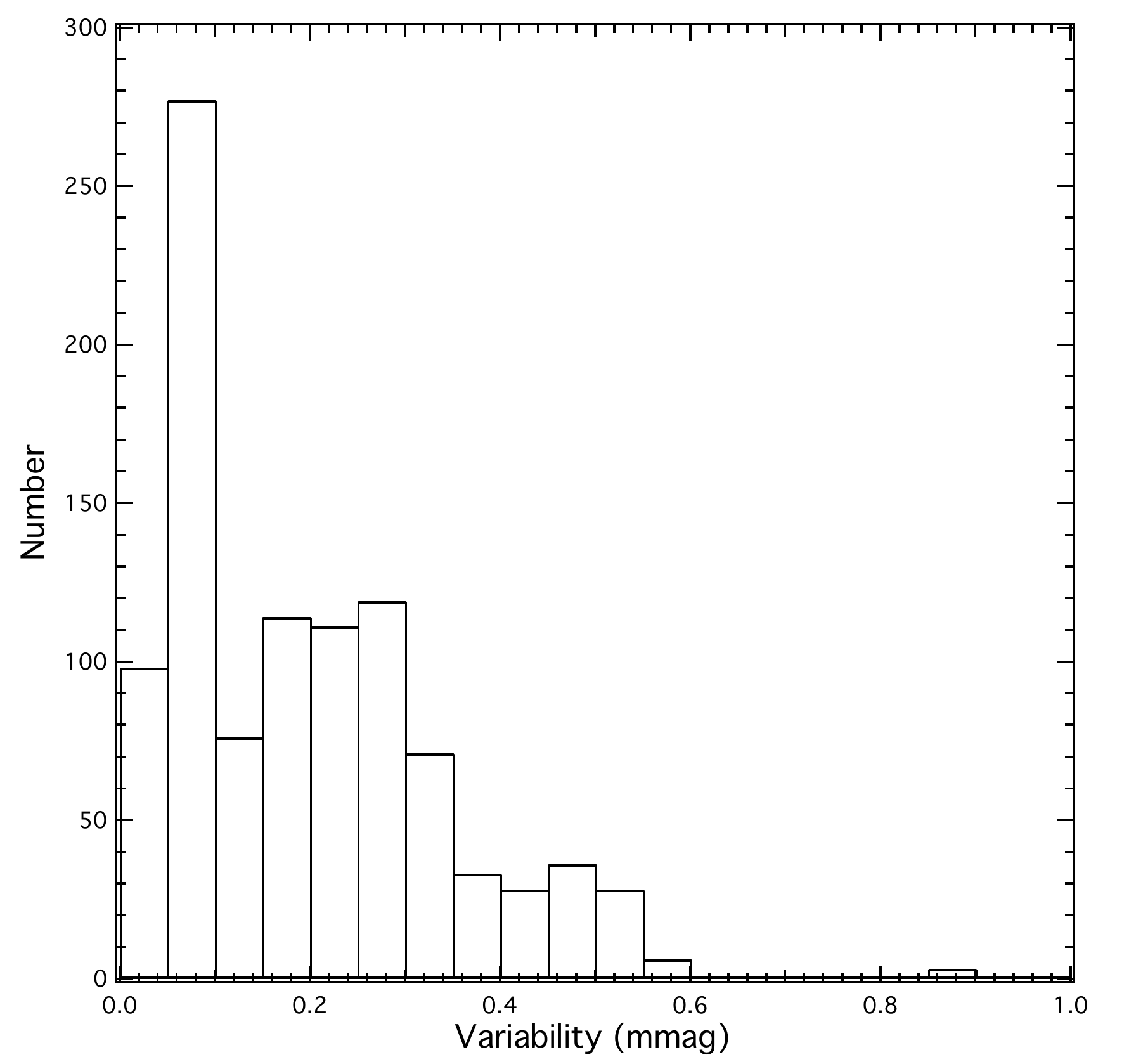}
\caption{Distribution of solar variability values calculated from the same dataset as Figure 6.}
\end{figure}

\begin{table}
\caption{Solar variability as cumulative percentages}
\label{obs}
\begin{tabular}{@{}lc}
\hline
Variability limit & percent \\
(mmag) & \\

\hline
$< 0.06$ & 16.1 \\
$< 0.07$ & 24.4 \\
$< 0.10$ & 37.5 \\
$< 0.13$ & 41.6 \\
$< 0.20$ & 56.5 \\
$< 0.30$ & 79.5 \\
$< 0.40$ & 89.9 \\
$< 0.50$ & 96.3 \\
$< 0.60$ & 99.7 \\
\hline
\end{tabular}

\end{table}

We list in Table 3 the percentage cumulative solar variability from our VIRGO samples. From this we find that nearly 42\% of the solar samples are smaller than the lowest corrected variability value among our M\,67 working sample stars.

\section{Data Analysis}
\subsection{Variables}

\citet{nard16} report the properties of 68 variables in the M\,67 field, 43 of which are new discoveries by them. The periods range from about 0.05 to nearly 39 days. Only 14 of these variable stars are present in our working sample of 639 stars. When we order our working sample stars from highest to lowest FVI, the first four entries are known variables, and 10 of the 14 other known variables in our sample are within the first 75 entries. This implies that many as yet undiscovered variables are lurking within our working sample.

We list in Table 4 the 14 known variables in our working sample. The table also includes the periods from \citet{nard16}, as well as our period determinations from the {\it Kepler} data and the periodogram category. There are multiple reasons for the discrepancies between the two sets of periods. First, it is not surprising that the agreement is poor for the stars in the ``mult'' category, since they do not have a single dominant period. As is evident from the two eclipsing binaries in the table, our periodogram analysis method derives periods for them that are exactly half the true period.

The first star on the list is also the faintest and has large estimated errors by \citet{nard16}, up to about one magnitude (but most errors they list for this star are much smaller)! The estimated errors for the {\it Kepler} fluxes are much smaller for the same magnitude star. For this star, the smallest estimated error in the $R$ lightcurve of \citet{nard16} is 0.05 magnitude (for the 180s exposures); the {\it Kepler} data have a mean estimated error near 6 mmag. In addition, the {\it Kepler} flux error values tend to vary little for a given star; such is not the case with the \citet{nard16} estimated photometric errors.

What's more, the ground-based data is sampled differently than the {\it Kepler} data. The data \citet{nard16} used to determine periods was obtained over a span of 764 days (with just under 30 nights of actual observation). The {\it Kepler} data is a nearly continuous series over a span of about 75 days. The {\it Kepler} data has the advantage in that it avoids problems with aliasing, but its shorter baseline limits the maximum measurable periods to about 40 days. While \citet{nard16} did determine white light magnitudes (a good match to the {\it Kepler} spectral response) in addition to $BVRI$ magnitudes, they suffer from greater photometric systematic error.

\begin{table}
\caption{Known variables in the M\,67 field in our working sample}
\label{obs}
\begin{tabular}{@{}lccccl}
\hline
ID\# & EPIC & $V$ & P1 & P2 & Cat \\
 &  &  & (d) & (d) & \\

\hline
N101 & 211391190 & 18.76 & 0.614410 & 0.153770 & mult \\
N193 & 211393067 & 16.88 & 5.014566 & 4.859178 & rot \\
N211 & 211393420 & 14.45 & 19.180715 & 13.033594 & rot \\
N236 & 211394170 & 12.43 & 2.863171 & 15.029691 & mult \\
N634 & 211400944 & 14.94 & 23.975894 & 13.839223 & mult \\
N1122 & 211407971 & 14.94 & 16.486412 & 14.533132 & mult \\
N1188 & 211408858 & 12.71 & 5.521103 & 6.526756 & mult \\
N1447 & 211412192 & 12.85 & 0.441437 & 0.220718 & mult \\
N1570 & 211413815 & 13.32 & 0.360466 & 0.180235 & EB \\
N1746 & 211416111 & 12.66 & 1.356193 & 1.358766 & mult \\
N1776 & 211416577 & 11.15 & 1.067112 & 0.533873 & EB \\
N2409 & 211427165 & 13.87 & 2.802946 & 2.820888 & rot \\
N2450 & 211427909 & 14.68 & 27.675604 & 15.484050 & mult \\
N2562 & 211430343 & (15.57) & 4.803519 & 22.153803 & mult \\
\hline
\end{tabular}

\medskip
Notes: The ID\#'s, $V$ magnitudes and P1 periods are from the \citet{nard16} catalog. P2 is the period determined in the present work from the {\it Kepler} data. For the last star, the $R$ magnitude is given.
\end{table}

\subsection{Single Versus Binary Variability}

How do the periods and variability compare among the different classes in our sample? In order to answer this question, we compare the distributions of the photometric periods of our sample stars in the ``SM'' and ``SN'' classes in panel {\bf a} of Figure 8. In panel {\bf b} of Figure 8 we compare the periods of the stars in the ``SM'' and ``BM'' classes. These figures include all the categories. A two-sample Kolmogorov-Smirnov (K-S) test of the samples in panel {\bf a} yields a p-value above 0.5, which means we cannot reject the hypothesis that the two samples are drawn from the same parent population. In the case of the two samples in panel {\bf b}, however, the p-value is only 0.008, implying different parent populations.

To determine the expected age distribution of non-member stars in the field of M\,67, we ran a Galactic simulation  using the online Besoncon web interface.\footnote{http://model.obs-besancon.fr/} The 1155 simulated stars cover the same range in $R$ magnitude as our working sample. The mean age for the simulated stars in the region of the M\,67 field is 6.3 Gyr. This is about 2 Gyr older than the age of M\,67. It is not surprising, then, that the field stars have a similar period distribution compared to the M\,67 members.

\begin{figure}
\includegraphics[width=3.5in]{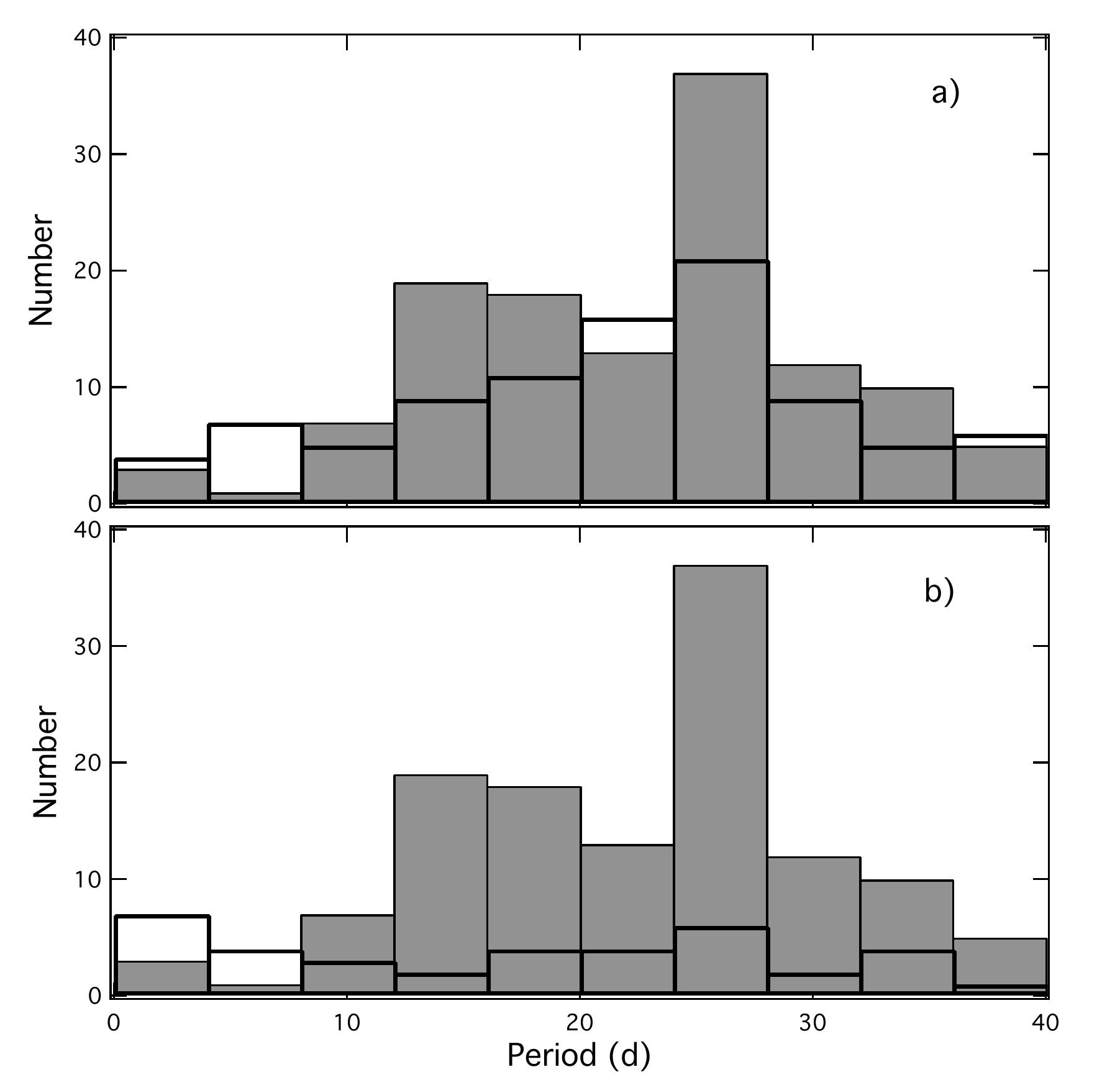}
\caption{Period distributions for stars in the ``SM'' (shaded bars) and ``SN'' (empty bars) classes are compared in panel {\bf a}. The ``SM'' (shaded bars) and ``BM'' (empty bars) classes are compared in panel {\bf b}.}
\end{figure}

In Figure 9 we compare the variability and period distributions of the stars in the ``SM'' and ``BM'' classes that are in the ``rot'' category. Only 45 ``SM'' stars and 14 ``BM'' stars are included in the plots. Applying K-S tests to the data in each panel of Figure 8, we find that they are likely drawn from the same distributions.

\begin{figure}
\includegraphics[width=3.5in]{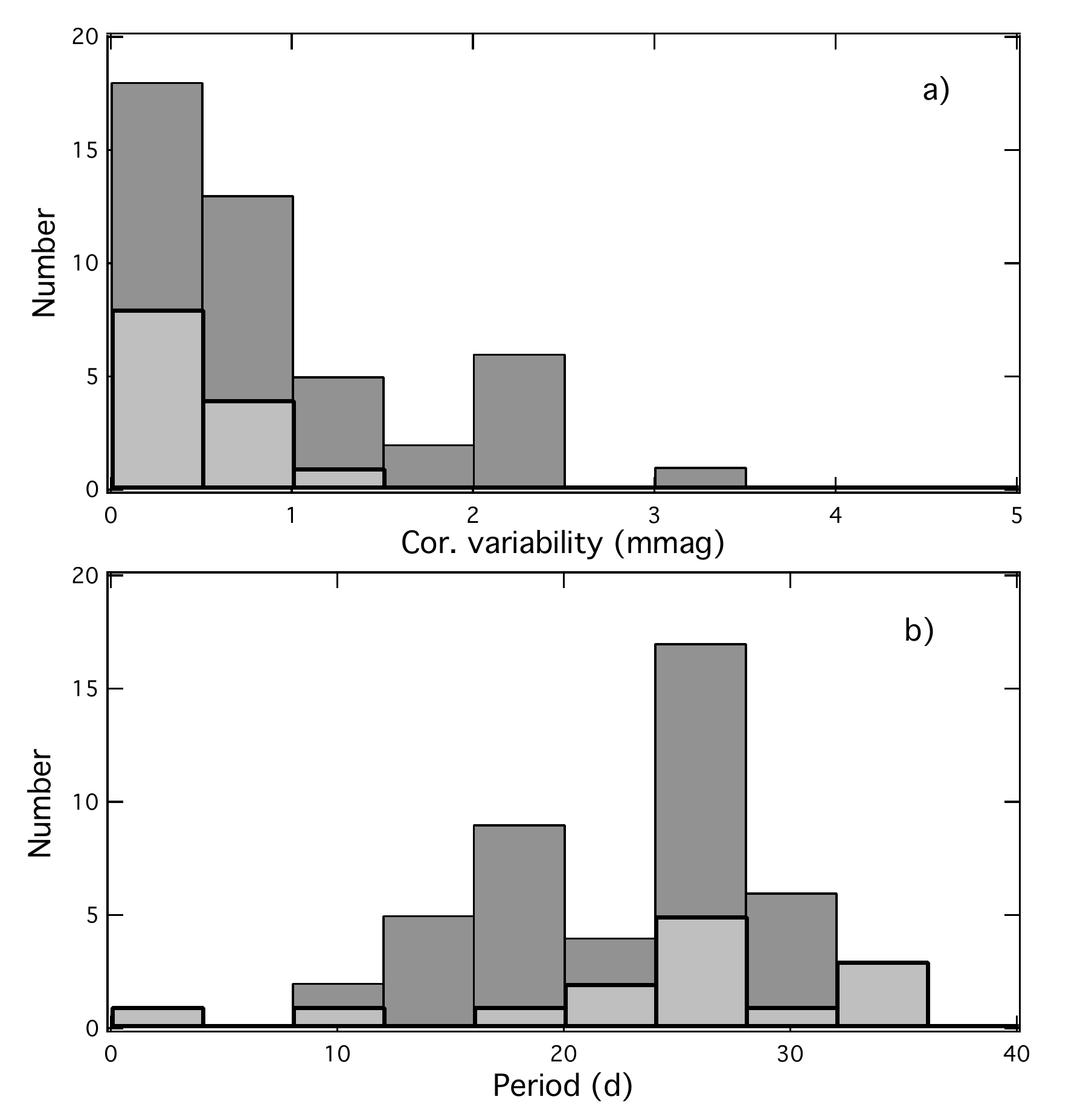}
\caption{Distributions of the corrected variability values for stars in the ``SM'' (dark shaded bars) and ``BM'' (light shaded bars) classes and within the ``rot'' category are compared in panel {\bf a}. The periods are compared for these sample samples in panel {\bf b}.}
\end{figure}

\subsection{Solar Analogs}

Several authors have identified solar analog  candidates in M\,67. The best such candidate is S770, which was examined in detail by \citet{onehag11} using high quality spectra (they actually identified it as a solar twin). \citet{gell15} identify ten stars in M\,67 stars as solar analogs; we list their properties in Table 5. Two stars in the table (S996, S1462) are actually binaries and should be removed from the list of solar analogs.  Only one star, S945, is also in our working sample. According to our periodogram analysis, it has a period of 25.9 days; its periodogram and phased light curve are show in Figure 10.

In order to identify additional solar analog candidates in M\,67, we compared the $V$ and $B-V$ photometry of the stars in \citet{nard16} (that are also included in the {\it Kepler} input catalog) to that of S770. A star must pass several tests before we can consider it as a candidate. First, a star must be within about 0.2 magnitudes in $V$ and about 0.12 magnitudes in $B-V$ from S770. Second, it must be a member of M\,67 and a single star, as indicated by \citet{gell15}; in other words, it must be in the class ``SM.'' In total, we identity an additional 32 solar analog candidates in M\,67, which we list in Table 6. The nine new candidates with {\it Kepler} light curves have period values that fall into two narrow ranges near 15 and 25 days, with the latter having twice as many stars as the first. The periods don't appear to have any relation with the periodogram category. The variability values average near 1.0 mmag. Of the new candidates listed in Table 5, S1602 is the most similar to S770 (Figure 11). 

\begin{figure}
\includegraphics[width=3.5in]{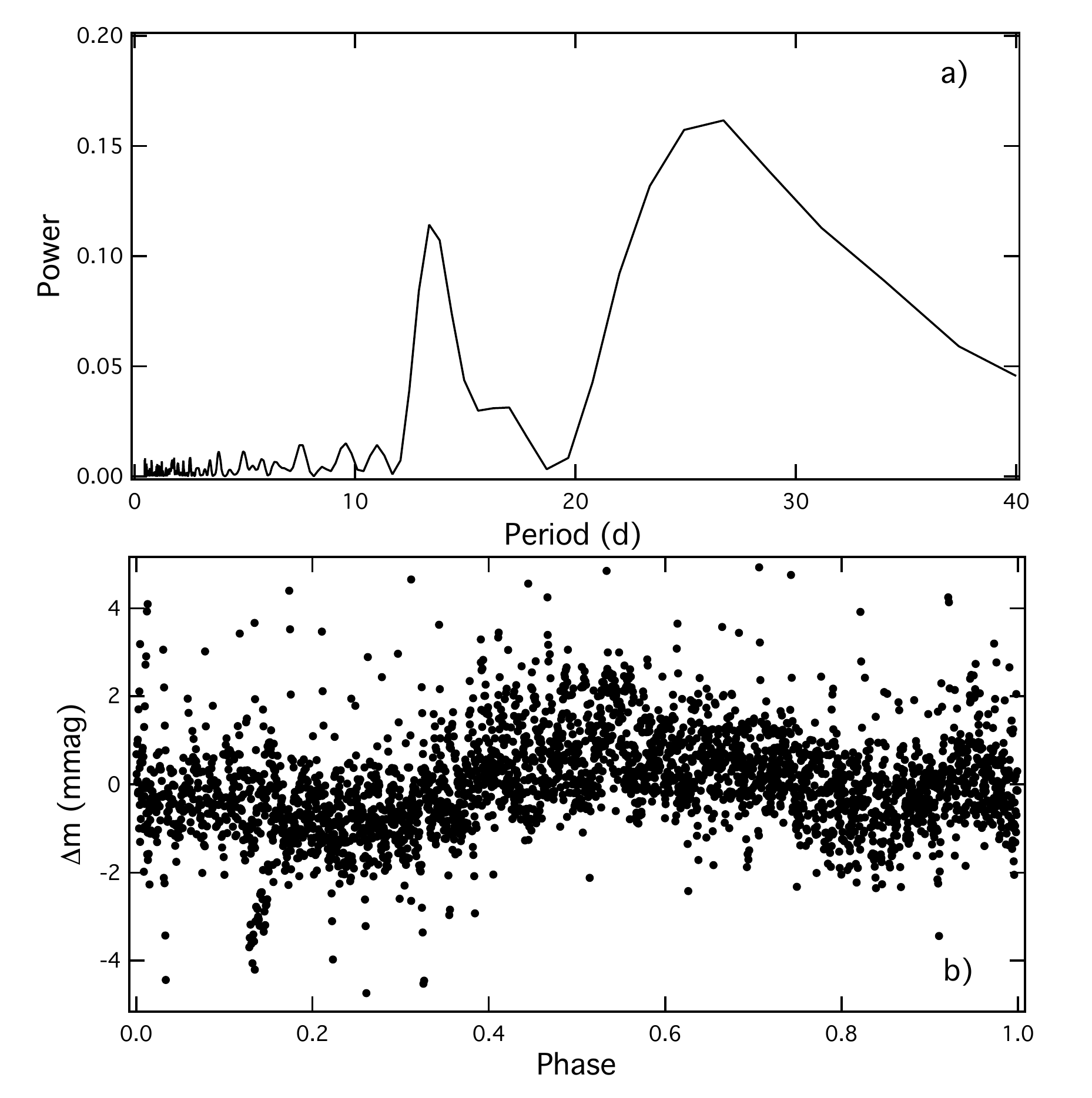}
\caption{Periodogram of S945 (panel {\bf a}). Phased light curve of S945 using the optimum period (25.9 d) extracted from the periodogram (panel {\bf b}).}
\end{figure}

\begin{figure}
\includegraphics[width=3.5in]{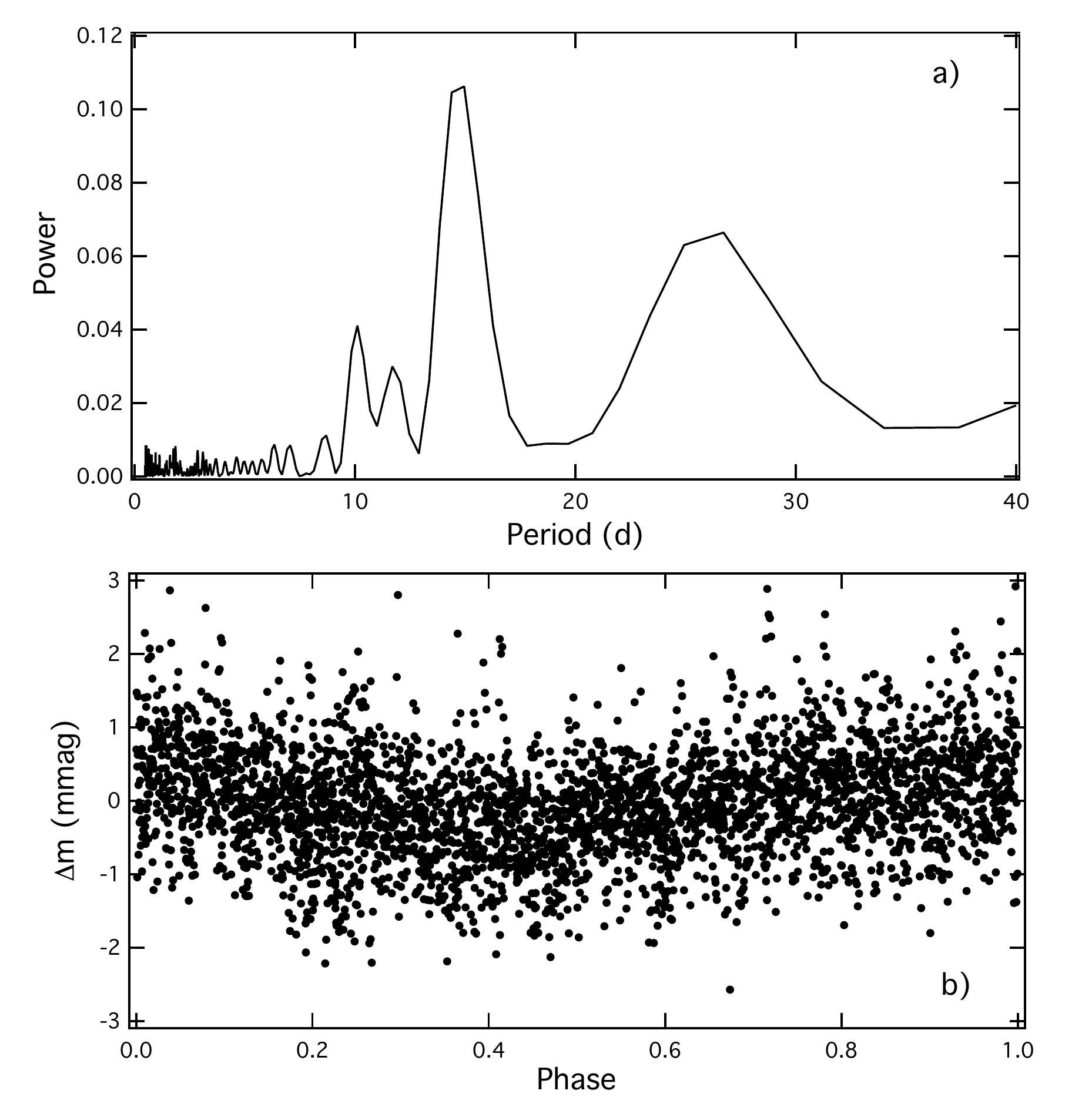}
\caption{Same as Figure 10 but for S1602.}
\end{figure}

\begin{table}
\caption{Stars in M\,67 identified as solar analogs by \citet{gell15}}
\label{obs}
\begin{tabular}{@{}rcccccl}
\hline
ID\# & EPIC & $V$ & $B-V$ & Variability & Class & Cat \\
 &  &  &  & (mmag) &  & \\
\hline
S770 & 211411531 & 14.625 & 0.731 & --- & SM & ---\\
S779 & 211412674 & 14.660 & 0.697 & --- & SM & ---\\
S785 & 211414090 & 14.847 & 0.716 & --- & SM & ---\\
S945 & 211400500 & 14.552 & 0.700 & 1.08 & SM & rot\\
S996 & 211409139 & 15.050 & 0.858 & --- & BM & ---\\
S1041 & 211412691 & 14.746 & 0.690 & --- & SM & ---\\
S1095 & 211420648 & 14.568 & 0.670 & --- & SM & ---\\
S1335 & 211425037 & 14.643 & 0.646 & --- & SM & ---\\
S1462 & 211412824 & 14.303 & 0.713 & --- & BM & ---\\
S2211 & 211410700 & 14.743 & 0.679 & --- & SM & ---\\
\hline
\end{tabular}

\medskip
Notes: The ID\#'s in this table and the next are from \citet{sand77}. The $V$ and $B-V$ photometry in this table and the next are from \citet{nard16}.
\end{table}

\begin{table}
\caption{Additional solar analog candidates in M\,67}
\label{obs}
\begin{tabular}{@{}rcccccl}
\hline
ID\# & EPIC & $V$ & $B-V$ & Variability & Period & Cat \\
 &  &  &  & (mmag) & (d) & \\
\hline
S491 & 211417658 & 14.818 & 0.733 & 1.20 & 23.5 & mult\\
S616 & 211410502 & 14.684 & 0.865 & --- & --- & ---\\
S622 & 211413212 & 14.523 & 0.659 & 0.62 & 24.3 & rot\\
S629 & 211417973 & 14.760 & 0.722 & --- & --- & ---\\
S724 & 211397319 & 14.560 & 0.671 & 0.73 & 24.4 & rot\\
S753 & 211407194 & 14.636 & 0.670 & --- & --- & ---\\
S777 & 211412436 & 14.565 & 0.660 & --- & --- & ---\\
S802 & 211416296 & 14.818 & 0.691 & --- & --- & ---\\
S928 & 211394644 & 14.807 & 0.732 & 0.97 & 15.0 & rot\\
S955 & 211402253 & 14.793 & 0.748 & --- & --- & ---\\
S965 & 211405030 & 14.727 & 0.753 & --- & --- & ---\\
S966 & 211405256 & 14.533 & 0.647 & --- & --- & ---\\
S991 & 211408535 & 14.558 & 0.710 & --- & --- & ---\\
S1106 & 211422726 & 14.750 & 0.698 & --- & --- & ---\\
S1184 & 211395699 & 14.688 & 0.704 & 2.31 & 23.8 & mult\\
S1204 & 211401708 & 14.640 & 0.820 & --- & --- & ---\\
S1218 & 211405832 & 14.604 & 0.683 & --- & --- & ---\\
S1258 & 211410536 & 14.517 & 0.654 & --- & --- & ---\\
S1341 & 211428580 & 14.811 & 0.656 & 1.38 & 25.6 & mult\\
S1421 & 211400746 & 14.507 & 0.695 & --- & --- & ---\\
S1422 & 211400915 & 14.629 & 0.698 & --- & --- & ---\\
S1430 & 211403620 & 14.718 & 0.756 & --- & --- & ---\\
S1477 & 211417575 & 14.591 & 0.694 & --- & --- & ---\\
S1481 & 211418998 & 14.776 & 0.751 & --- & --- & ---\\
S1484 & 211419792 & 14.508 & 0.840 & --- & --- & ---\\
S1602 & 211411477 & 14.648 & 0.730 & 0.67 & 14.7 & rot\\
S1616 & 211416103 & 14.510 & 0.679 & --- & --- & ---\\
S1621 & 211420635 & 14.529 & 0.692 & 0.75 & 27.4 & rot\\
S1714 & 211411621 & 14.642 & 0.885 & 1.24 & 15.2 & mult\\
S1724 & 211418075 & 14.615 & 0.751 & --- & --- & ---\\
S1729 & 211421134 & 14.781 & 0.793 & --- & --- & ---\\
S1806 & 211402217 & 14.768 & 0.798 & --- & --- & ---\\
\hline
\end{tabular}

\medskip
\end{table}

\section{Discussion}

Photometric variation due to starspots declines with age. For this reason, ground-based photometric measurements are limited to clusters younger than about 1 Gyr. A major motivation to measure the rotation periods of stars in old clusters is to extend the empirical calibration of the gyro-age method to older stars. The gyro-ages appear to be inconsistent with isochrone ages for older stars, and there also appears to be a difference between cluster and field stars \citep{kov15}. 

M\,67 is the oldest cluster for which stellar rotation periods have been attempted to be measured. The prior record holder was the 2.5 Gyr old NGC 6819, for which \citet{bal13} and \citet{mei15} measured the rotation periods from the {\it Kepler} main mission data. \citet{mei15} determined a mean rotation period of 18.2 d for Sun-like stars in the cluster from data spanning 3.75 years. The period values in their Figure 2 show very little scatter at a given color. In contrast, our results for the Sun-like stars in M\,67 (Figure 9) show large scatter at a given color. There are several reasons for this difference.

First, the stars in M\,67 are older, making it more difficult to detect photometric variation due to starspots. \citet{mei15} were only able to measure rotation periods for a fraction of the dwarfs in NGC 6819 (see their Extended Data Figure 6); furthermore, of the 43 stars with detected rotation periods, they eventually retained only 30 stars. In contrast, we derived periods for nearly every star in our sample, even those with weak variations. Second, our M\,67 data only spans just under 80 days, which is only enough to see about three full rotation periods for solar-age stars. Finally, even if all the dwarf stars in M\,67 at a given color have the same rotation period, our numerical periodogram experiments with the solar irradiance fluxes (Figure 5) show that we should expect to obtain a period distribution with standard deviation near eight days.

From their measured rotation periods and the model of \citet{barn10}, \citet{mei15} derived a gyro-age of 2.49 Gyr for NGC 6819. This is consistent with independent age estimates \citep{jeff13}. We follow the approach of Meibom et al. to derive a gyro-age for M\,67. We begin by compiling a list of the single member stars in the ``rot'' category (plotted in Figure 12). The mean rotation period of the 28 stars in the figure with $(B-V)_{\rm 0}$ values between 0.55 and 0.90 magnitudes (the same color range used by Meibom et al. for NGC 6819) is $23.4 \pm 6.5$ d ($\pm 1.2$ d standard error of the mean). From this we derive a gyro-age estimate of $3.7 \pm 0.3$ Gyr for M\,67. This age is consistent with the age range estimated by \citet{sara09} for M\,67 of 3.5 to 4.0 Gyr from stellar isochrones.

\begin{figure}
\includegraphics[width=3.5in]{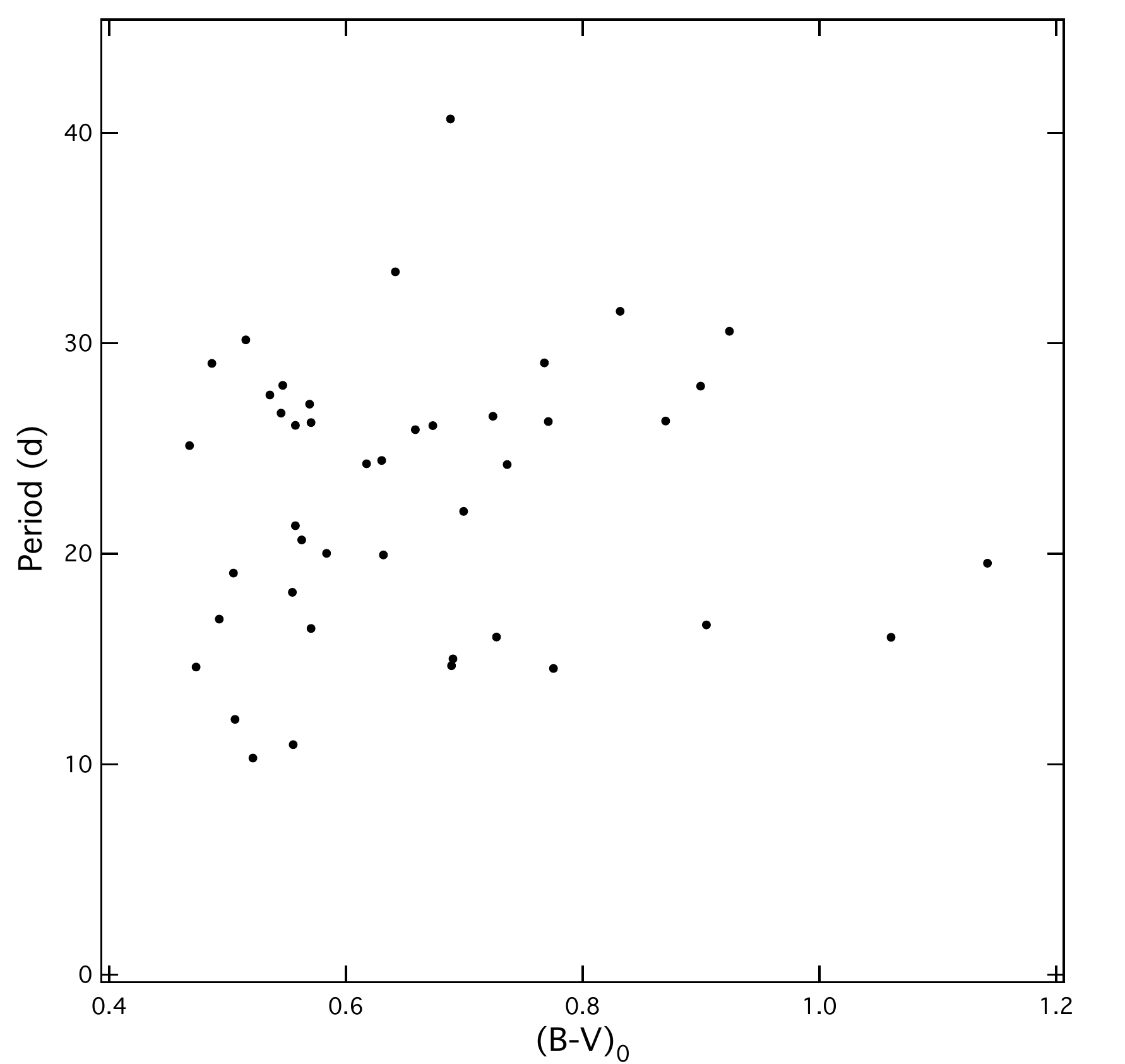}
\caption{Period as a function of dereddened color for stars in the ``SM'' class and in the ``rot'' category. The $V$ and $(B-V)$ photometry used to prepare the plot are from \citet{nard16}, and the assumed color excess is 0.041 magnitude \citep{tay07}. The $V$ and $K_{\rm p}$ magnitudes of these stars average near 14.7 and 14.6, respectively. Note, two stars have been left out of the plot because they lack color data.}
\end{figure}

Our comparison of the photometric variability of stars in M\,67 to that of the Sun is similar to the study of \citet{basri13}. We both compare solar irradiance VIRGO data to {\it Kepler} data of Sun-like stars; they use stars in the main {\it Kepler} mission field, while we use {\it K2} M\,67 data. Our study, however, has two advantages. First, the time baseline of the VIRGO data we employed is significantly longer (1996-2015 versus 1996-2009), better sampling the range of solar activity. Second, stars in the field of the main {\it Kepler} mission do not have well-constrained ages. This contrasts with our M\,67 member stars sample, which is homogeneous and well-constrained in age. Our studies also differ in how they handle the measurement uncertainties. \citet{basri13} attempt to model the measurement errors with a parameterized noise equation, calibrated with the quietest stars in their sample. In contrast, we assume the {\it K2} estimated error data products reliably report the measurement errors and that systematic errors have been properly accounted for in the Bayesian pipeline analysis.

\section{Conclusions}

We examined the light curves of stars from the {\it Kepler/K2-Campaign-5} that are included in the M\,67 field. We calculated variabilities and derived periods for 639 stars. The mean period of the Sun-like single cluster members with the better quality period determinations is 23.4 days. This implies an age near 3.7 Gyr, assuming the photometric periods are due to rotational modulation. This is consistent with recent age determinations of M\,67 based on stellar evolution models.

The intrinsic photometric variability of solar analogs in M\,67 is greater than the solar variability measured from the VIRGO fluxes. This is consistent with M\,67 being younger than the Sun, but the difference seems too large to explain in terms of age difference alone. The variabilities of many stars in M\,67 are such that they could be detected from the ground with extensive photometric observations.

\section*{Acknowledgments}

We thank the anonymous reviewer for many helpful comments and suggestions. We acknowledge receipt of the updated dataset (version: 6\_004\_1510) of the VIRGO Experiment, on the cooperative ESA/NASA Mission SoHO from the VIRGO Team through PMOD/WRC, Davos, Switzerland.

\bsp

\label{lastpage}

\end{document}